\documentstyle[preprint,aps,amssymb,eqsecnum]{revtex}

\newtheorem{thm}{Theorem}

\newcommand{\SL}{\mathop{\mathrm{SL}}\nolimits}
\newcommand{\Sl}{\mathop{\mathrm{sl}}\nolimits}
\newcommand{\Tr}{\mathop{\mathrm{Tr}}\nolimits}
\newcommand{\diff}{\mathop{\mathrm{diff}}\nolimits}

\newcommand{\boldsymbol}[1]{\mbox{\boldmath$#1$}}

\newenvironment{proof}{\par
  \normalfont \topsep6pt plus6pt \trivlist
  \item[\hskip\labelsep\itshape Proof.]\ignorespaces}{%
  \hspace*{\fill}$\square$\endtrivlist}

\begin{document}

\preprint{ICN-UNAM 99-03 Mexico; CRM-2606 Montreal}

\title{Solutions of Nonlinear Differential and Difference Equations
  with Superposition Formulas}

\author{Alexander Turbiner\cite{leave}${}^{\dagger}$}
\address{Instituto de Ciencias Nucleares, UNAM\\ Apartado Postal 70--543,
04510 Mexico D.F., Mexico\\
\texttt{turbiner@xochitl.nuclecu.unam.mx}}

\author{Pavel Winternitz}
\address{Centre de recherches math\'ematiques, Universit\'e de
  Montr\'eal\\ Montr\'eal, Quebec, Canada, H3C 3J7, Canada\\
\texttt{wintern@crm.umontreal.ca}}

\date{\today}

\maketitle

\begin{abstract}
Matrix Riccati equations and other nonlinear ordinary
differential equations with superposition formulas are, in the
case of constant coefficients, shown to have the same exact
solutions as their group theoretical discretizations. Explicit solutions of certain classes of scalar and matrix Riccati equations are presented as
an illustration of the general results.
\end{abstract}

\pacs{}

\section{Introduction}

The concept of linear superposition of solutions of a linear
equation, the Schr\"odinger equation, forms the basis of all of
quantum mechanics. It has two different aspects. One is that a
linear combination of two solutions is again a solution. The
second is that the general solution of a linear equation can be
written as a linear combination of a sufficient number of
particular solutions.

Both of these aspects of superposition have their counterpart in
the theory of nonlinear phenomena, as represented by nonlinear
differential and difference equations. Thus, in soliton theory,
specific solutions, namely solitons, can be ``superposed'' in a
nonlinear manner, to obtain multisoliton solutions of integrable
nonlinear partial differential equations\cite{13}. On the other
hand, certain classes of nonlinear ordinary equations exist, that
allow a nonlinear superposition formula, in the sense that their
general solution can be expressed analytically in terms of an
appropriate number of particular solutions.

These two different aspects of ``nonlinear superposition'' are
actually related via B\"acklund transformations.

Several recent articles have been devoted to a study of nonlinear
ordinary differential equations (ODEs)  with superposition
formulas \cite{1,2,3,4,5,6}.  The entire approach is based on a theorem due to
Sophus Lie \cite{7} that states the following.

\begin{thm}[S. Lie]
The general solution $\roarrow
y(t;c_1,\ldots,c_n)$ of a system of $n$ ordinary differential
equations
\begin{equation}
\dot{\roarrow{y}} = \roarrow{\eta}(\roarrow{y}, t),\quad  \text{$\roarrow{y} \in
  {\mathbb C}^n$ (or ${\mathbb R}^n$)}\label{1.1}
\end{equation}
can be written as a function of $m$ particular solutions and $n$
arbitrary constants
\begin{equation}
\roarrow{y} = \roarrow{F}
\bigl(\roarrow{y}_1(t),\ldots,\roarrow{y}_m(t); c_1,\ldots,c_n)\bigr)
\label{1.2}
\end{equation}
if  and only if:
\begin{enumerate}
\item The system (\ref{1.1}) has the form
\begin{equation}
\dot{\roarrow{y}}(t) = \sum_{k=1}^r Z_k(t) \roarrow{\eta}_k(\roarrow{y},
t), \quad 1\leq r < \infty\label{1.3}
\end{equation}
\item The vector fields
\begin{equation}
\hat X_k = \sum_{\mu=1}^n \eta_k^\mu(t) \partial_{y_\mu},\label{1.4}
\end{equation}
generate a Lie algebra $L$ of finite dimension $\dim L = r
<\infty$. The number of equations $n$, number of fundamental
solutions $m$ and the dimension of the Lie algebra $r$ satisfy
\begin{equation}
  nm \geq  r.\label{1.5}
\end{equation}
\end{enumerate}
\end{thm}

A geometric approach to such equations has been developed
\cite{5}. The variables $y_1$,\dots, $y_n$ are considered to be
coordinates on a homogeneous space $M \sim G/G_0$ with the group
$G$ acting transitively on $M$. The Lie algebra $L$ of vector
fields (\ref{1.4}) is the Lie algebra corresponding to the group
$G$. The subgroup $G_0$ is the isotropy group of the origin and
its Lie algebra $L_0$ is the subalgebra of $L$ consisting of
vector fields that vanish at the origin. Thus, it is possible to
associate a system of ODEs satisfying Lie's criterion with every
group-subgroup pair $(G_0,G)$, or with every Lie algebra pair
$(L_0,L)$, $L_0 \subset L$.

We call the system (\ref{1.3}) decomposable if a subsystem of
equations can be split off that involves only a subset of
variables, namely, $y_1$,\dots, $y_k$, $1\leq k < n$, that itself
satisfies Lie's theorem. This occurs if the action of $G$ on $M$
is not primitive, i.e., if a $G$-invariant foliation of $M$
exists \cite{5} Indecomposable systems of ODE's with
superposition formulas are obtained if the pair of Lie algebras
$(L_0, L)$ forms a transitive primitive Lie algebra
\cite{8,9,10,11,12}. This requires that $L_0$ be a maximal
subalgebra of $L$, not containing any ideal of $L$. Transitive
primitive Lie algebras over ${\mathbb C}$ and ${\mathbb R}$ have
been classified \cite{8,9,10,11,12}. Let us now consider a
particularly interesting class of transitive primitive Lie
algebras, namely one where $L$ is a simple classical Lie algebra
over ${\mathbb C}$ or ${\mathbb R}$ and $L_0$ is a maximal
parabolic subalgebra. The space $M
\sim G/G_0$ is then a Grassmanian. Using affine coordinates
on this space we obtain vector fields with polynomial
coefficients of order at most four \cite{5}. The corresponding
ODEs (\ref{1.3}) have the same polynomial nonlinearities.

These equations have the following properties:
\begin{enumerate}
\item By construction, they allow a nonlinear superposition
formula.

\item They are linearizable by introducing homogeneous coordinates
on the Grassmannian, i.e., by an embedding into a larger system of
linear equations.

\item They have the Painlev\'e property \cite{13}, i.e., their
singularity structure is particularly simple: the only movable
singularities of their solutions are poles.
\end{enumerate}

A discretization of the nonlinear ODEs with superposition
formulas has been proposed in a recent article \cite{14}. It
preserves all of these properties,  in particular the Painlev\'e
property, manifesting itself as singularity confinement
\cite{15}.

The discretization was performed for the case when $L$ was any
classical complex Lie algebra and $L_0$ any of its maximal
parabolic subalgebras. The discretization made use of the
linearization of the equations (\ref{1.3}). It can, however, be
presented schematically without using the corresponding linear
system.

Indeed, let (\ref{1.4}) represent the vector fields of Lie's
theorem. Eq.~(\ref{1.3}) can be rewritten as
\begin{eqnarray}
\dot{\roarrow{y}} &=& \widehat{X} (t) \roarrow{y}\label{1.6}\\
\widehat{X}(t) &=& \sum_{k=1}^r Z_k(t)\sum_{\mu=1}^n \eta_k^\mu(t)
\partial_{y_\mu},\label{1.7}
\end{eqnarray}

As $t$ varies, the vector field $\hat X (t)$ follows some path in
the Lie algebra $L$. The vector fields (\ref{1.4}) can be
integrated to give the corresponding local group action
\begin{equation}
\roarrow{y} =g  \roarrow{u},\label{1.8}
\end{equation}
where $\roarrow{u}$ is some chosen point on $M$. The form of
action (\ref{1.8}) is independent of the equation (\ref{1.6}),
i.e. of the functions $Z_k(t)$. It involves $r = \dim L = \dim G$
group parameters. The general solution of eq.~(\ref{1.6}) can be
represented in the form (\ref{1.8}), where $g = g(t)$ is a
function of $t$. As $t$ varies, $g(t)$ will trace a path in the
group manifold, corresponding to the path that $\widehat{X}(t)$
follows in the Lie algebra. Reconstructing this path is
equivalent to solving the equation~(\ref{1.6}).

The difference equations with all the desired properties are
obtained in the form (\ref{1.8}) with $g$  a function of time (i.e.,
each of the $r$ group parameters in  $g$ depends on time). The
equations are
\begin{equation}
\label{1.9}
\bar y = g y,\quad y\equiv y(t),\quad \bar y\equiv y(t+\sigma),
\end{equation}
where $\sigma$ is the spacing of the lattice. The group element
$g = g(t)$ is to be evaluated at the same point as $y(t)$ on the
right hand side. For $\sigma \to 0$ we have $g (t) = I + \sigma
\widehat{X}(t)$ and the difference equation (\ref{1.9}) goes into
the differential equation (\ref{1.6}).

The purpose of this article is to explore some properties of the
difference equations (\ref{1.9}) and in particular to show that
in the case of constant coefficients in the differential
equations (\ref{1.6}) the solutions of the continuous and
discrete equations coincide exactly.

A general theorem on solutions of continuous and discrete matrix
Riccati equation is formulated and proven in Section~\ref{sec2}.
Section~\ref{sec3} is devoted to examples.

\section{Continuous and Discrete Matrix Riccati Equations}
\label{sec2}

Let us consider the Grassmannian  $M \sim G/G_0$ where $G$
is $\SL(N,{\mathbb C})$ and $G_0$ is a maximal parabolic subgroup. We have
\begin{equation}
\label{2.1}
N=p+q,\quad G \sim
\left( \begin{array}{cc}
G_{11} & G_{12} \\ G_{21} & G_{22}
\end{array}  \right),\quad
\ G_0 \sim
\left( \begin{array}{cc}
G_{11} & 0 \\ G_{21} & G_{22}
\end{array}  \right),
\end{equation}
where $G_{11} \in {\mathbb C}^{p\times p}$, $G_{12} \in {\mathbb
C}^{p\times q}$,$G_{21} \in {\mathbb C}^{q\times p}$, $G_{22} \in
{\mathbb C}^{q\times q}$ and $M = pq$, $\det G =1$. Affine
coordinates on this space (in the neighborhood of the origin) are
given by the matrix elements of a matrix $W \in {\mathbb
C}^{p\times q}$. The nonlinear ODEs with superposition formulas
in this case are the Matrix Riccati Equations \cite{3,4,5,6,16}
\begin{equation}
\label{2.2}
\dot W = A + BW + WC + WDW,
\end{equation}
where
\begin{equation}
\label{2.3}
\xi (t) =
\left( \begin{array}{cc}
B(t) & A(t) \\ -D(t) & -C(t)
\end{array}  \right),\quad \Tr B =\ Tr C,
\end{equation}

For $t$ fixed, $\xi(t)$ is an element of the Lie algebra
$\Sl(N,{\mathbb C})$. The division into blocks is the same in eq.
(\ref{2.3}) as in (\ref{2.1}) The matrices $A$, $B$, $C$, $D$ are
known functions of $t$. As $t$ varies the matrix $\xi(t)$ follows
some path lying in $\Sl(N,{\mathbb C})$.

The Matrix Riccati Equation (\ref{2.2}) is linearized by introducing
homogeneous coordinates, i.e. the matrix elements of two matrices
$X$ and $Y$. We put
\begin{equation}
\label{2.4}
W = X Y^{-1},\quad X \in {\mathbb C}^{p\times q},\quad
Y \in {\mathbb C}^{q\times q}.
\end{equation}
The origin on $M$ corresponds to $X=O$, $Y = I$.   When the
matrices $X$, $Y$ satisfy the linear system
\begin{equation}
\label{2.5}
\left( \begin{array}{c}
\dot X \\ \dot Y
\end{array}  \right) =
\left( \begin{array}{cc}
B(t) & A(t) \\ -D(t) & -C(t)
\end{array}  \right)
\left( \begin{array}{c}
X \\ Y
\end{array}  \right)
\end{equation}
the matrix $W = XY^{-1}$  satisfies eq.~(\ref{2.2}).

The ordinary Riccati equation for $w(t)$ a scalar function, namely
\begin{equation}
\label{2.6}
\dot w = a + (b+c)w + dw^2 ,
\end{equation}
is obtained for $p=q=1$. For $p\geq 2$, $q=1$ eq.~(\ref{2.2}) is
called the \emph{projective Riccati} equation, since it
corresponds to the projective action of $\SL(p+1, {\mathbb C})$
on the space ${\mathbb C}^p$.

The discretization of eq.~(\ref{2.5}) is \cite{14}
\begin{equation}
\label{2.7}
\left( \begin{array}{c}
\overline{X} \\ \overline{Y}
\end{array}  \right) =
\left( \begin{array}{cc}
B(t) & A(t) \\ -D(t) & -C(t)
\end{array}  \right)
\left( \begin{array}{c}
X \\ Y
\end{array}  \right)
\end{equation}
where $\overline{X}= X(t+\sigma)$, $\overline{Y}= Y(t+\sigma)$ and the discrete
matrix Riccati equation is obtained via the projection (\ref{2.4}),
i.e.,
\begin{equation}
\label{2.8}
\overline{W} = (G_{11}W+ G_{12}) (G_{21}W+G_{22})^{-1}.
\end{equation}

Now let us restrict ourselves to the case of  the Matrix Riccati
Equations  (\ref{2.2}) with constant coefficients. In this case
the matrix $\xi$ of eq.~(\ref{2.3}) can be viewed as a
one-dimensional subalgebra of $\Sl(N,{\mathbb C})$ and the
corresponding one-dimensional subgroup of $\SL(N,{\mathbb C})$
is obtained by exponentiating $\xi$
\begin{equation}
\label{2.9}
G = e^{\epsilon \xi} = I + \epsilon \xi +
\frac{1}{2!}\epsilon^2 \xi^2 + \cdots.
\end{equation}
This matrix $G$ is to be inserted into eq.~(\ref{2.7}) and
eq.~(\ref{2.8}), first identifying the group parameter $\epsilon$
with the lattice spacing $\sigma$, for example, putting
$\epsilon=\sigma$. Alternatively the form of the transformation
$G$ can be obtained by integrating the vector fields on the right
hand side of eq.~(\ref{2.2}). To do this we rewrite
eq.~(\ref{2.2}) in components as
\begin{equation}
\label{2.10}
\dot W_{\alpha a} = A_{\alpha a} + B_{\alpha \beta}W_{\beta a} +
W_{\alpha b}C_{b a} + W_{\alpha b}D_{b \beta}W_{\beta a},
\end{equation}
where $1 \leq \alpha \leq p$, $1\leq a \leq q$ (repeated Greek
labels are summed from 1 to $p$, Latin ones from 1 to $q$).
Eq.~(\ref{2.8}) is obtained directly by integrating
\begin{equation}
\label{2.11}
\frac{d\overline{W}_{\alpha a}}{d\epsilon} = A_{\alpha a} +
B_{\alpha \beta}\overline{W}_{\beta a} + \overline{W}_{\alpha b}C_{b a} +
\overline{W}_{\alpha b}D_{b \beta}\overline{W}_{\beta a},
\end{equation}
where $\overline{W}_{\alpha a}|_{\epsilon=0} = W_{\alpha a}$.

We now  come to our main result which we formulate as a theorem.

\begin{thm}
The general solution of the continuous matrix Riccati equation
(\ref{2.2}) with constant coefficients coincides with the general
solution of the discrete matrix Riccati equation (\ref{2.8}),
with coefficients $G_{\mu \nu}$ satisfying eq.~(\ref{2.9}).
\end{thm}

\begin{proof}
Let us rewrite the discrete linear system (\ref{2.7}) as
\begin{equation}
\label{2.12}
e^{\sigma d/dt} Z = e^{\sigma \xi} Z,\quad Z = \left( \begin{array}{c}
X\\ Y\end{array}  \right)
\end{equation}
and expand both sides in powers of $\sigma$:
\[
1+\sigma \frac{d}{dt} + \frac{\sigma^2}{2} \biggl( \frac{d}{dt}
\biggr)^2 +\cdots+ \frac{\sigma^n}{n!} \biggl( \frac{d}{dt}
\biggl)^n +\cdots = 1+ \sigma\xi+\frac{\sigma^2}{2} \xi^2+ \cdots+
\frac{\sigma^n}{n!} \xi^n +\cdots.
\]
Comparing coefficients of the powers of $\sigma$ we have
\begin{equation}
\label{2.13}
\frac{d^n}{dt^n}Z = \xi^n Z.
\end{equation}
For $n = 0$ this is satisfied trivially, for $n=1$ we obtain the
continuous linear system (\ref{2.5})
\begin{equation}\label{2.14}
  \dot Z = \xi Z.
\end{equation}
Let $Z(t)$ be a solution of eq.~(\ref{2.14}) and hence $W$ a solution of the
continuous matrix Riccati equation (\ref{2.2}). In order for $Z$ to be a
solution of the discrete system (\ref{2.7}), and hence $W$ a solution of the
discrete matrix Riccati equation (\ref{2.8}), $Z$ must also satisfy
eq.~(\ref{2.13}) for all $n\geq 2$. For $n = 2$ we have
\begin{equation}
\label{2.15}
\frac{d^2 Z}{dt^2}=\frac{d}{dt}(\xi Z) =
\frac{d\xi}{dt}Z+\xi\frac{dZ}{dt}= \xi^2 Z,
\end{equation}
since $d\xi/dt=0$ and $\dot Z=\xi Z$. For $n\geq 3$
eq.~(\ref{2.13}) follows by induction, hence every solution of eq.~(\ref{2.14})
with constant $\xi$ satisfies the difference equation~(\ref{2.7}).
\end{proof}

The general solution of eq.~(\ref{2.5}) depends on $p.q$
significant free parameters, the initial conditions. The same is
true for the difference equation (\ref{2.7}).  Hence, the general
solution of the differential equation~(\ref{2.5}) provides the
general solution of  the difference equation~(\ref{2.7})  and
vice versa.

\section{Examples}\label{sec3}

\subsection{The Scalar Riccati Equation}

Let us first consider a scalar Riccati equation with real
constant coefficients:
\begin{equation}\label{3.1}
  \dot u= a+bu+cu^2,\quad c\neq 0 ,
\end{equation}
and introduce $\Delta \equiv b^2- 4ac$. We simplify this equation by
using an appropriate linear transformation $u=\alpha y+\beta$.
Depending on the sign of $\Delta$ we obtain three different standard
forms of (\ref{3.1}).

\subsubsection{$\Delta > 0$}

We choose $\alpha$ and $\beta$  so as to obtain
\begin{equation}
\label{3.2}
  \dot y = \omega(y^2 - 1),\quad \omega= \mathrm{const}.
\end{equation}
Integrating the vector field $\hat X=(y^2+1)d/dy$ we
obtain the \emph{discrete} Riccati equation
\begin{equation}\label{3.3}
\bar y= \frac{y \cosh \sigma\omega - \sinh \sigma\omega}{-y \sinh \sigma\omega +
\cosh \sigma\omega},
\end{equation}
where $y=y(t)$, $\bar y=y(t+\sigma)$. It is easy to verify by direct
calculation that the general solution of both equations
(\ref{3.2}), (\ref{3.3}) is
\begin{equation}\label{3.4}
  y= - \coth \omega(t-t_0),
\end{equation}
where $t_0$ is an arbitrary constant.

\subsubsection{$\Delta < 0$}

An appropriate choice of $\alpha$ and $\beta$ yields
\begin{equation}
\label{3.5}
  \dot y = \omega(y^2 + 1),\quad \omega= \mathrm{const},
\end{equation}
Integration of the vector field
$\widehat{X}=(y^2-1)d/dy$ provides us with the discrete Riccati
equation
\begin{equation}
\label{3.6}
\bar y= \frac{y \cos \sigma\omega + \sin \sigma\omega}{-y \sin \sigma\omega +
\cos \sigma\omega}.
\end{equation}
The general solution of these equations is
\begin{equation}
\label{3.7}
  y= \tan \omega(t-t_0),
\end{equation}
where $t_0$ is an arbitrary constant.

\subsubsection{$\Delta = 0$}

In this case the Riccati equation (\ref{3.1}) can be
reduced to
\begin{equation}\label{3.8}
  \dot y = y^2,
\end{equation}
and its discrete form is
\begin{equation}
\label{3.9}
\bar y = \frac{y}{1-\sigma y}.
\end{equation}
The general solution of both
equations (\ref{3.8}) and (\ref{3.9}) is
\begin{equation}
\label{3.10}
  y=- \frac{1}{t-t_0},
\end{equation}
where $t_0$ is an arbitrary constant.

\subsection{Coupled Riccati Equations Based on a Nonprimitive
Group Action}\label{sec3.2}

Most of the earlier articles on nonlinear ODEs with
superposition formulas concerned indecomposable systems of
equations. A study of imprimitive actions and hence decomposable
systems was initiated quite recently \cite{17}.

Here we shall give an example of such a system and its
discretization. Consider the Lie algebra $\Sl(2,{\mathbb R})$, realized as a
subalgebra of $\diff(2,{\mathbb R})$, by the vector fields.
\begin{equation}
\label{3.11}
X_1=\partial_x,\quad X_2=x\partial_x+y\partial_y,
\quad X_3=x^2\partial_x+(2xy+ky^2)\partial_y,
\end{equation}
when $k=0$, or $k=1$.

The corresponding ODEs are
\begin{equation}\label{3.12}
  \dot x= a+bx+cx^2,\quad
  \dot y= by+c(2xy+ky^2),
\end{equation}
Let us now restrict to the case when $a$, $b$, $c$ are constants and
moreover assume that $b^2-4ac < 0$, $c \neq 0$. We can then, by means
of a linear transformation of $x$ take eq.~(\ref{3.12}) into the standard form
\begin{equation}\label{3.13}
  \dot x=\omega(x^2+1),\ \dot y=\omega(2xy+ky^2),\quad \omega=\mathrm{const}.
\end{equation}
To obtain a discretization that preserves the underlying group
structure, we integrate the vector field
\begin{equation}
\label{3.14}
\widehat{X}=\omega[(1+x^2)\partial_x+(2xy+ky^2)\partial_y],
\end{equation}
We have
\begin{equation}
\label{3.15}
\frac{d\bar x}{d\sigma}=\omega(1+ \bar x^2),\quad
\frac{d\bar y}{d\sigma}=\omega(2\bar x \bar y + k\bar y^2),
\end{equation}
where $\bar x|_{\sigma=0}=x$, $\bar y|_{\sigma=0}=y$. The result is
\begin{eqnarray}
\bar x &=& \frac{x\cos \sigma\omega + \sin \sigma\omega}{-x \sin \sigma\omega +
\cos \sigma\omega},\nonumber\\
\label{3.16}
\bar y &=& \frac{y(x^2+1)}{D},
\end{eqnarray}
where $D=(x^2+1+kxy)(-x \sin \sigma\omega +
\cos \sigma\omega)^2-ky (x \cos \sigma\omega + \sin \sigma\omega)
(-x\sin \sigma\omega + \cos \sigma\omega)$ Thus, eq.~(\ref{3.16})
is a difference equation that has eq.~(\ref{3.13}) as its
continuous limit (for $\sigma \to 0$).

Eq.~(\ref{3.13}) is easy to solve and we obtain
\begin{eqnarray}
x &=& \tan \omega (t-t_0),\nonumber\\
\label{3.17}
y &=& \frac{1}{\alpha \cos^2 \omega(t-t_0)- k[\sin 2\omega(t-t_0)]/2},
\end{eqnarray}
where $\alpha$ and $t_0$ are integration constants. Expression
(\ref{3.17}) is also the general solution of the difference equation
(\ref{3.16}).

\subsection{Projective Riccati Equations}

Let us now consider the matrix Riccati equation (\ref{2.2}) for
$q=1$ which is called the \emph{projective} Riccati equations.

We rewrite the system as
\begin{equation}\label{3.18}
\dot{\mathbf{y}} = \mathbf{a} +  B \mathbf{y} + \mathbf{y}^{\mathsf{T}} C \mathbf{y}
\end{equation}
where $\mathbf{a}$, $\mathbf{y} \in K^p$ are column vectors, $B$,
$C \in K^{p\times p}$ are matrices and $\mathsf{T}$ denotes
transposition. We consider the case when all coefficients are
constants.

We simplify eq.~(\ref{3.18}) by a linear transformation with constant
coefficients
\begin{equation}
\label{3.19}
  \mathbf{y} = A \mathbf{u} + \boldsymbol{\rho}
\end{equation}
Choosing $A$ and $\rho$ appropriately we can, for instance,
transform $C$ and $B$ in such a way that
\begin{equation}\label{3.20}
C_i=\delta_{i1},\ B_{j1}=0,\quad j=1,2,\ldots, p
\end{equation}
with some further freedom remaining.

Now let us consider the simplest case of $p = 2$.  Changing
notations we write the transformed projective Riccati equation
as
\begin{equation}
\label{3.21}
\dot x = 1 + y +x ^2,\quad \dot y = a+ by + xy,
\end{equation}
where $a$, $b$ are constants.

In order to discretize the equations we must either exponentiate
a matrix as in eq.~(\ref{2.9}) or integrate the vector field
\begin{equation}
\label{3.22}
\widehat{X} = (1+y+x^2) \partial_x + (a+by+xy)\partial_y,
\end{equation}
We choose the second procedure and put
\begin{equation}
\label{3.23}
\frac{d{\bar x}}{d\sigma}= (1+\bar y + \bar x^2),\quad
\frac{d{\bar y}}{d\sigma}= (a+ b \bar y +\bar x \bar y),
\end{equation}
where $\bar x|_{\sigma=0}=x$, $\bar y|_{\sigma=0}=y$.

Eliminating $\bar y$ from eq.~(\ref{3.23}) we have
\begin{equation}
\label{3.24}
\ddot {\bar x}=3 \bar x \dot{\bar x}+b\dot{\bar x}- \bar x ^3-b \bar x^2-\bar x + a - b
\end{equation}
where dots denote differentiation with respect to $\sigma$.

Eq.~(\ref{3.24}) has the Painlev\'e property and can be transformed to
one of the 50 standard forms \cite{18}, studied by Painlev\'e and
Gambier. More specifically it is equivalent to the linearizable
equation No.~VI in the list reproduced by Ince \cite{18}  (see p.~334). Indeed, we put
\begin{equation}
\label{3.25}
\bar x(\sigma)=\alpha(\sigma) w(\tau) + \beta,\quad \tau=\tau(\sigma),
\end{equation}
while the constant $\beta$ must satisfy:
\[
\beta^3+b\beta^2 + \beta + b -a =0
\]
The functions $\alpha(\sigma)$ and $\tau(\sigma)$ must satisfy linear
equations with constant coefficients:
\begin{eqnarray}
\label{3.26}
&&\ddot \alpha - (b+3\beta)\dot \alpha + (3\beta^2+2b\beta+1)\alpha =0\\
\label{3.27}
&&\dot \tau = - \alpha.
\end{eqnarray}
The function $w(\tau)$ then satisfies
\begin{eqnarray}
\label{3.28}
  w_{\tau\tau} &=& -3w w_\tau-w^3+q(\tau)(w_\tau+w^2)\\
\label{3.29}
  q(\tau) &=&\frac{3\dot \alpha-3\alpha\beta-b\alpha}{\alpha^2}.
\end{eqnarray}
Equation (\ref{3.28}) is linearized by the Cole-Hopf transformation
\begin{equation}
\label{3.30}
  w=\frac{f_{\tau}}{f},\quad
  f_{\tau\tau\tau}=q(\tau)f_{\tau\tau}.
\end{equation}
Eq.~(\ref{3.30}) can be solved in quadratures, however, we shall only
consider the simplest case, when $q(\tau)=0$ . This implies a
constraint between  the constants $a$, $b$, namely
\begin{equation}
\label{3.31}
  q(\tau)=0,\quad a=\frac{2b(b^2+9)}{27}
\end{equation}
Equation (\ref{3.30}) in this case  is solved trivially. Returning to
the original variables $\bar x$ and $\bar y$ we have
\begin{eqnarray}
x(\sigma)&=&-k\frac{2Pe^{k\sigma}+Q}{Pe^{k\sigma}+Q+Re^{-k\sigma}}+\beta
\nonumber\\
\label{3.32}
y(\sigma)&=&k\frac{4(\beta -k)Pe^{k\sigma}+(2\beta -k)Q}{Pe^{k\sigma}+Q+Re^{-k\sigma}}
-\beta^2-1.
\end{eqnarray}
Here $P$, $Q$, $R$ are integration constants and we have put
\begin{eqnarray}
k=\beta+\frac{b}{3}\neq 0,&\quad&
\beta=\frac{-b+\sqrt{3(b^2-3)}}{3},\ b^2\neq 3,\nonumber\\
\label{3.33}
\tau=e^{k\sigma},&& \alpha=-ke^{k\sigma}
\end{eqnarray}
We eliminate the constants $P,Q,R$ by imposing initial conditions
as in eq.~(\ref{3.23}). Finally, we obtain difference equations
\begin{equation}\label{3.34}
\bar x=\frac{g_{11}x+g_{12}y+g_{13}}{g_{31}x+g_{32}y+g_{33}},\quad
\bar y=\frac{g_{21}x+g_{22}y+g_{23}}{g_{31}x+g_{32}y+g_{33}},
\end{equation}
where
\begin{eqnarray}
g_{11}&=&(\beta -2k)(2\beta -k)e^{k\sigma}-4(\beta -k)^2+
\beta(2\beta -3k)e^{-k\sigma},
\nonumber\\
g_{12} &=& (\beta -2k)e^{k\sigma}+2(k- \beta)+\beta e^{-k\sigma},
\nonumber\\
g_{13} &=& (\beta -2k)(-\beta^2+ k\beta +1)e^{k\sigma}-2(\beta -k)+\beta e^{-k\sigma}
\nonumber\\
g_{21} &=& (4k\beta -4k^2-\beta^2-1)(2\beta -k)e^{k\sigma} -
4(2\beta k-k^2-\beta^2-1)(\beta -k)\nonumber \\
&&\quad {}- (\beta^2+1)(2\beta -3k)e^{-k\sigma},
\nonumber\\
g_{22} &=& (4k\beta -4k^2-\beta^2-1)e^{k\sigma} -2
(2\beta k-k^2-\beta^2-1)-(\beta^2+1)e^{-k\sigma},
\nonumber\\
g_{23} &=& (4k\beta -4k^2-\beta^2-1)(-\beta^2+\beta k+1)e^{k\sigma} -
2(2\beta k-k^2-\beta^2-1)(2\beta k+1)\nonumber \\
&& \quad {}+
(\beta^2+1)(\beta^2+2k^2-3\beta k+1)e^{-k\sigma},
\nonumber\\
g_{31} &=& (2\beta -k)e^{k\sigma} - 4(\beta -k) + (2\beta - 3k) e^{-k\sigma},
\nonumber\\
  g_{32} &=& e^{k\sigma}-2 + e^{-k\sigma},
\nonumber\\
g_{33} &=& (1+\beta k-\beta^2)e^{k\sigma}+2(\beta^2-2\beta k+1)+
(-\beta^2 -2k^2 + 3\beta k +1) e^{-k\sigma}.\label{3.35}
\end{eqnarray}
As in the previous examples the general solutions of the
projective Riccati equations (\ref{3.21}), and their
discretization (\ref{3.34}), coincide.

\section{Conclusion}

The results of this article can be summed up as follows.
\begin{enumerate}
\item We
have shown that matrix Riccati equations with constant
coefficients and their group theoretical discretizations have the
same solutions.
\item We have shown how other types of equations
with superposition formulas can be discretized via an integration
of the underlying vector fields. For constant coefficients the
continuous and discrete equations again have the same solutions.
\end{enumerate}
An open question, presently under study, is whether the results
of this article can be used to obtain an effective  perturbative
approach to solving differential and difference equations that
allow superposition formulas, but do not have constant
coefficients.

\acknowledgments

The work reported here was started while P.~W. was visiting
ICN-UNAM in Mexico, D.F. and finished while he was visiting the
Centro Internacional de Ciencias in Cuernavaca, Mexico. He thanks
both institutions for their hospitality. The research of P.~W. is
supported in part by research grants from NSERC of Canada and
FCAR du Qu\'ebec.

\end{document}